\documentclass[journal=jctcce,manuscript=article]{achemso}
\usepackage{graphicx}
\usepackage{amsmath}
\usepackage{amsfonts}
\usepackage{amssymb}
\usepackage{amsthm}
\usepackage{braket}
\usepackage{subfigure}
\usepackage{multirow}
\usepackage[utf8]{inputenc}
\usepackage{bm}
\usepackage{comment}
\newlength\tindent
\newcommand{\parz}[2]{ \frac{\partial{#1}}{\partial{#2}}}            

\author{Tommaso Giovannini$^{\top}$}
\affiliation{Department of Chemistry, Norwegian University of Science and Technology, 7491 Trondheim, Norway}
\author{Luca Bonatti$^{\top}$}
\affiliation{Scuola Normale Superiore,
             Piazza dei Cavalieri 7, 56126 Pisa, Italy.}
\author{Marco Polini}
\affiliation{Dipartimento di Fisica dell'Universit\'a di Pisa, Largo Bruno Pontecorvo 3, I-56127 Pisa, Italy}
\alsoaffiliation{Istituto  Italiano  di  Tecnologia,  Graphene  Labs,  Via  Morego  30,  16163  Genova,  Italy}
\author{Chiara Cappelli}
\affiliation{Scuola Normale Superiore,
             Piazza dei Cavalieri 7, 56126 Pisa, Italy.}
\email{chiara.cappelli@sns.it}

\graphicspath{{img/}} 

\title[]
 {Graphene Plasmonics: a Novel Fully Atomistic Approach for Realistic Structures}

\usepackage{booktabs}                                  
\usepackage{units}                                     
\usepackage{cancel}                                    
\usepackage{braket}
\newcommand*{\EF}{\text{E}_\text{F}}

\begin{document}

\begin{center}
$^{\top}$ T.G. and L.B. contributed equally to this work.
\end{center}

\begin{abstract}
We demonstrate that the plasmonic properties of realistic graphene and graphene-based materials can effectively and accurately be modeled by a novel, fully atomistic, yet classical, approach, named $\omega$FQ. Such model is able to reproduce all plasmonic features of these materials, and their dependence on shape, dimension and fundamental physical parameters (Fermi energy, relaxation time and two-dimensional electron density). Remarkably, $\omega$FQ is able to accurately reproduce experimental data for realistic structures of hundreds of nanometers ($\sim$ 370.000 atoms), which cannot be afforded by any \emph{ab-initio} method. Also, the atomistic nature of $\omega$FQ permits the investigation of complex shapes, which can hardly be dealt with by exploiting widespread continuum approaches.
\end{abstract}

\newpage

\section*{Introduction}

Graphene \cite{geim2010rise} has emerged as an outstanding plasmonic material \cite{grigorenko2012graphene,garcia2014graphene} because it provides a strong field confinement with relatively low losses \cite{woessner2015highly} that cannot be reached by noble metal plasmons \cite{johnson1972optical}. 
Additionally, graphene plasmons can be easily tuned by exploiting electrical gating \cite{ju2011graphene,fei2012gate,
fei2011infrared,yan2012tunable,fang2013gated,fang2013active,brar2013highly,
chen2012optical,shin2011control}. Such an unique property, together with the possibility of chemical doping, provides an easy mechanism to tune the Fermi energy in graphene\cite{chen2011controlling} and consequently the Plasmon Resonance Frequency (PRF). In this context, an accurate modeling of the optical response of both homogeneous and nanopatterned graphene is of crucial relevance in the development of devices or in the understanding of complex physical phenomena, as for instance Graphene Enhanced Raman Scattering (GERS).\cite{ling2009can,ling2010first,ling2013graphene,ling2012charge}

The optical properties of graphene have been so far computationally investigated  \cite{koppens2011graphene,jablan2009plasmonics,vakil2011transformation,thongrattanasiri2012quantum,
fang2013gated,cox2016quantum,christensen2011graphene,gruneis2008tight} by exploiting continuum classical approaches (e.g. the Boundary Element Method - BEM) \cite{de2002retarded,koppens2011graphene,vakil2011transformation} or \emph{ab-initio} methods.\cite{reich2002tight,jablan2009plasmonics,yan2011nonlocal}

Continuum models have low computational cost,\cite{fang2013gated} but they lack any atomistic description of the 2D material and therefore cannot describe finite-size, edge effects \cite{thongrattanasiri2012quantum} and defects. The latter can indeed be treated by \emph{ab-initio} methods\cite{cox2016quantum,thongrattanasiri2012quantum,brey2006electronic,brey2007elementary,han2007energy}, but at an high computational cost, which hampers the study of large, realistic structures.

In this work, we present a novel, fully atomistic, yet classical, approach (named $\omega$FQ) able to reproduce all plasmonic features of graphene and graphene-based materials. Remarkably, our method overcomes most of the limitations of current continuum and \emph{ab-initio} methods; in fact, it gives results quantitatively comparable to \emph{ab-initio} but for large, realistic structures of more than $\sim$ 370.000 atoms, which can normally be tackled only with continuum models, because they are completely not affordable by \emph{ab-initio} approaches. Large systems can in principle be described by means of continuum approaches, however in case of complex shapes, basic electrodynamical continuum methods cannot be applied and numerical methodologies (such BEM) able to treat complex boundaries need to be exploited.\cite{bonatti2020frontiers} Their use is far from being trivial, thus limiting their application to realistic systems. 


\section*{Results and discussion}

$\omega$FQ finds its theoretical foundations in the fact that the plasmonic response of 2D carbon-based materials is dominated by the synchronous excitation of $\pi$ electrons, which is mediated by the electrodynamical conductance. In this framework, in $\omega$FQ each graphene carbon atom is endowed with a complex, electric charge $q_i$, whose value is not fixed but varies as a response to the external oscillating electric field. The classical equation of motion which specifies the charges is obtained by modeling the charge exchange as regulated by the Drude model, which mimics the electrodynamical conductance. The charge flow between atoms then occurs because of the difference in their chemical potential.\cite{giovannini2019classical}\\
In the frequency domain, the classical $\omega$FQ equation of motion for charges reads as following (see Electronic Supplementary Information - ESI for the complete derivation):

\begin{equation} 
-i\omega q_{i} = \dfrac{2\tau v_\text{F}}{1- i \omega \tau} \sqrt{\frac{n_{2D}}{\pi}}\sum_{j} f(l_{ij}) \cdot \dfrac{A_{ij}}{l_{ij}} \cdot (\mu_{j}^{el} - \mu_{i}^{el})
\label{eq:wFQ}
\end{equation}

where, $\omega$ is the frequency of the external electric field, $\tau$ is the relaxation time, $v_\text{F}$ is the Fermi velocity (fixed to 10$^{6}$ m/s) and $n_{2D}$ is the 2D-density of graphene. $A_{ij}$ is the effective area connecting the \emph{i-th} and \emph{j-th} atoms, $l_{ij}$ is their distance and $\mu^{el}_i$ is the electrochemical potential of atom $i$. The Fermi energy is defined as $\EF = \hbar v_\text{F} \sqrt{\pi \cdot n_{2D}}$, whereas the effective mass $m^* = \sqrt{\pi\cdot n_{2D}}/v_\text{F}$.\cite{neto2009electronic} Finally, $f(l_{ij})$ is a function that guarantees that charge exchange only occurs between nearest neighbor atoms.
As it can be evinced by eq. \ref{eq:wFQ}, $\omega$FQ finds its strengths in the simplicity of the formulation and in the fact that the different parameters entering eq.\ref{eq:wFQ} can be directly recovered from experimental and/or computed data (see Table S1 in the ESI), thus potentially allowing its extension to 2D materials other than carbon-based. In addition, its fully atomistic nature permits to treat structural defects and/or chemical doping by simply modifying the input geometrical structures/parameters. 

In this work, the potentialities and the performance of $\omega$FQ are shown for four challenging, differently shaped graphene-based materials (nanoribbons, nanotriangles, nanodisks and nanorings). All the studied structures are planar and have been constructed by fixing the carbon-carbon distance (1.42 \AA).\cite{neto2009electronic} Notice that the 2D-density ($n_{2D}$) depends on the area of the material, therefore purely geometrical differences on the target systems directly reflect on the definition of $n_{2D}$ and indirectly on atom-atom couplings. 

\begin{figure}[htbp!]
	\centering
	\includegraphics[width=1\linewidth]{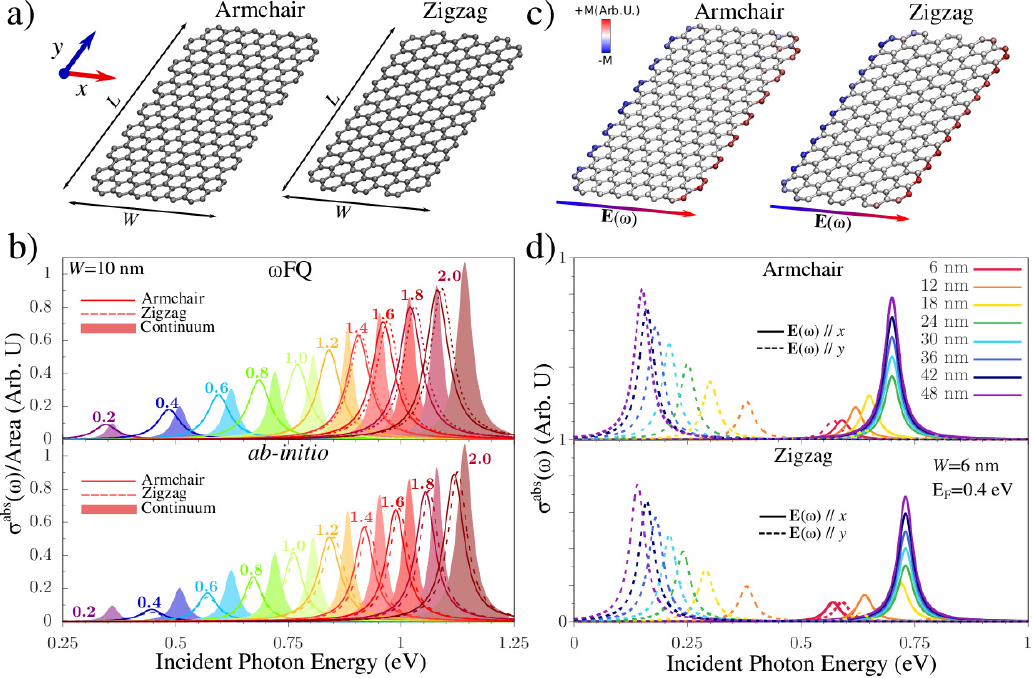}
	\caption{\textbf{(a)} Graphical depiction of the armchair and zigzag graphene nanoribbons studied in the present work. The two characteristic length scales $W$ (short edge) and $L$ (long edge) are highlighted. \textbf{(b)} $\omega$FQ (top) and \emph{ab-initio} $\sigma^{\text{abs}}$ of a graphene nanoribbon ($W$ = 10 nm and $L$ = 12 nm) as a function of the Fermi energy - $\EF$ (from 0.2 to 2.0 eV, with a constant step of 0.2 eV). Classical continuum results are also reported. Both Classical continuum and \emph{ab-initio} data are reproduced from Ref. \citenum{cox2016quantum}. In all cases, $x$-polarization is considered. \textbf{(c)} Pictorial representation of $\omega$FQ imaginary charges representing the local plasmonic response for AC and ZZ nanoribbons. Colors are satured for $\pm$ 3.0$^{-1}$ a.u. The external electric field intensity is 10$^{-7}$ a.u. \textbf{(d)} $\omega$FQ $\sigma^{\text{abs}}$ of armchair (top) and zigzag (bottom) graphene nanoribbons ($W$ = 6 nm) as a function of $L$ (from 6 nm (squared graphene sheet) to 48 nm, with a constant step of 6 nm). Both \textit{x}- (solid line) and \textit{y}- (dashed line) polarizations are considered. E$_{\text{F}}$ is 0.4 eV in all calculations.}
	\label{fig:ribbon}
\end{figure}

\begin{figure}[htbp!]
	\centering
	\includegraphics[width=1\linewidth]{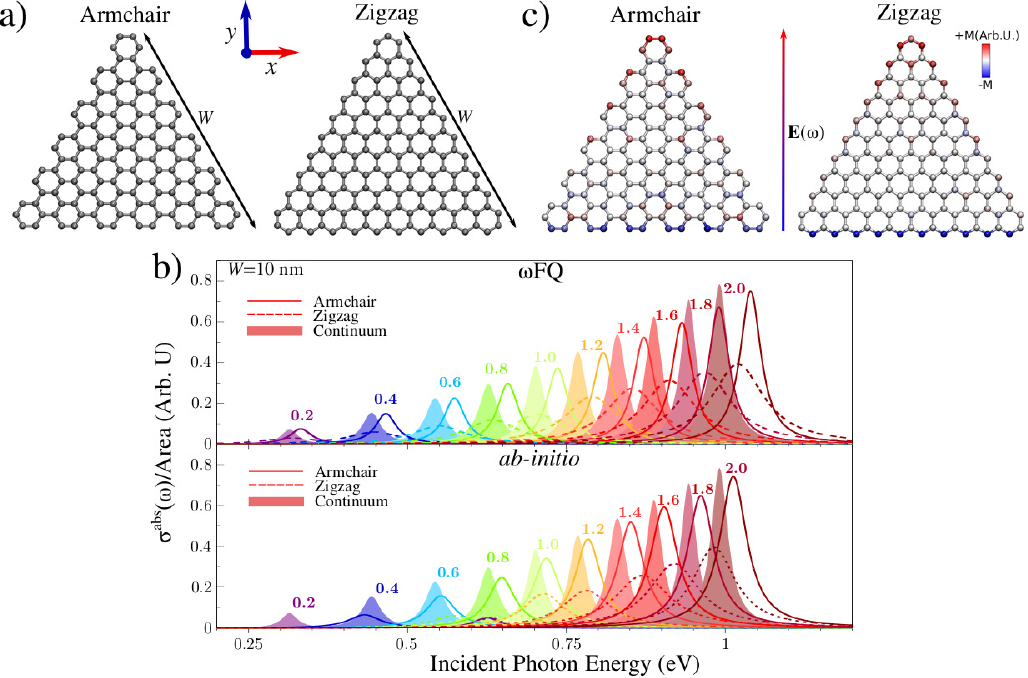}
	\caption{\textbf{(a)} Graphical depiction of the armchair and zigzag graphene nanotriangles studied in the present work. The main length scale (edge length) $W$ is highlighted. \textbf{(b)} $\omega$FQ (top) and \emph{ab-initio} $\sigma^{\text{abs}}$ of a graphene nanotriangle ($W$ = 10 nm) as a function of the Fermi energy - $\EF$ (from 0.2 to 2.0 eV, with a constant step of 0.2 eV). Classical continuum results are also reported. Both Classical continuum and \emph{ab-initio} data are reproduced from Ref. \citenum{cox2016quantum}. In all cases, $y$-polarization is considered. \textbf{(c)} Pictorial representation of $\omega$FQ imaginary charges representing the local plasmonic response for AC and ZZ nanotriangles. Colors are satured for $\pm$ 3.0$^{-1}$ a.u. The external electric field intensity is 10$^{-7}$ a.u.}
	\label{fig:triangles}
\end{figure}

We first show the performance of $\omega$FQ as applied to the optical response of graphene-based nanoribbons and nanotriangles, in two possible edge configurations, namely armchair-AC and zigzag-ZZ (see Figs. \ref{fig:ribbon} and \ref{fig:triangles}, panel a). Note that the atomistic nature of $\omega$FQ allows to discriminate among the two geometrical arrangements. 
The absorption cross section ($\sigma^{\text{abs}}$) of AC and ZZ nanoribbons with $W$ = 10 nm and $L$ = 12 nm (see Fig. \ref{fig:ribbon}a for definition) was computed as a function of the Fermi energy ($\EF$, see Fig. \ref{fig:ribbon}b): modifications in $\EF$ are widely exploited experimentally to tune the plasmonic response of graphene-based materials. Remarkably, due to its physical relevance, $\EF$ enters the definition of $\omega$FQ response equations (see Methods section).  $\omega$FQ absorption cross sections are compared with both classical-continuum and atomistic \emph{ab-initio} descriptions of the graphene-sheet, taken from Ref. \citenum{cox2016quantum}. 

The absorption cross section calculated by exploiting $\omega$FQ shows a prominent plasmon band which rapidly decreases in intensity and redshifts as $\EF$ decreases from 2.0 eV to 0.2 eV (see Fig. \ref{fig:ribbon}b, top). The nature of the plasmon responsible for the observed peak was investigated by plotting the imaginary $\omega$FQ charges calculated at the PRF for both AC and ZZ configurations (see Fig. \ref{fig:ribbon}c): clearly, the peak is associated with a dipolar plasmon. By deepening in Fig. \ref{fig:ribbon}b, $\omega$FQ calculations are in almost perfect agreement with \emph{ab-initio} data for both structural arrangements,\cite{cox2016quantum} provided that $\EF$ is greater than PRF. Under such condition, $\omega$FQ correctly reproduces the small differences between AC and ZZ edges, which are predicted \emph{ab-initio}, and this is particularly evident for $\EF > 1.0$ eV. This is indeed impressive, and highlights the capabilities of our fully atomistic method to accurately describe edge effects on the plasmonic response. Remarkably, the continuum approach cannot distinguish between AC and ZZ configurations due to its intrinsic limitations, which hamper a proper description of edge effects. 
The major discrepancies between $\omega$FQ and \emph{ab-initio} are reported if PRF $> \EF$: however, this does not influence the overall qualitative behavior of computed $\omega$FQ results. 

As stated before, one of the main features of graphene-based materials is the possibility to tune their optical response by modifying structural and electronic properties. Fig. \ref{fig:ribbon}d  reports calculated $\omega$FQ absorption cross sections of AC and ZZ graphene nanoribbons with $W$ = 6 nm, as a function of both the length of the sheet ($6 \leq L \leq 48$ nm) and the direction of the external electric field (either $x-$ or $y-$ polarizations). In the case of $x-$polarization (see Fig. \ref{fig:ribbon}d, solid lines) the PRF remains almost constant if the aspect ratio (i.e. $L/W$) is above 4 ($L$ = 24 nm, green line), independently of the considered configuration. On the contrary, PRF redshifts for smaller structures. The same does not apply to $y-$polarization (see Fig. \ref{fig:ribbon}d, dashed lines), for which PRF shifts to lower energies with respect to $x-$polarization. Also, PRF redshifts as the length of the nanoribbon increases. Such a behavior is perfectly in line with what is expected for an infinite nanoribbon with fixed width ($W$), which is characterized by a propagating plasmon. Therefore, the effects we are pointing out for $y-$polarization are entirely due to finite size effects. We notice also that, although computed imaginary polarizabilities are almost three times larger for $y-$polarization (as it is expected because the longer path of the dipolar plasmon lies on $y$ direction, see Fig. S4 in the ESI), $\sigma^{\text{abs}}$ is almost identical for the two perpendicular polarizations, due to its definition in terms of the external frequency (see Eq. \ref{eq:css_sec} in Methods). Therefore, a change in the field polarization allows for an alternative mechanism to tune the PRF of graphene nanoribbons. To the best of our knowledge, this aspect has not been attentively investigated in the previous literature.\cite{yan2013damping} 

We now move to discuss graphene-based nanotriangles (see Fig. \ref{fig:triangles}) in both AC and ZZ configurations (see Fig. \ref{fig:triangles}a, in which the main dimension $W$ is also highlighted).
In Fig. \ref{fig:triangles}b, calculated $\omega$FQ absorption corss sections are reported for a nanotriangle with $W$ = 10 nm as a function of $\EF$. \emph{Ab-initio} and continuum reference calculations, reproduced from Ref. \citenum{cox2016quantum}, are also shown for a direct comparison. Similarly to the previous case, the $\omega$FQ absorption cross section is dominated by a band which rapidly redshifts and decreases in intensity by decreasing $\EF$. Contrary to nanoribbons, large differences are obtained between AC and ZZ band widths, which reflect the different scattering times ($\tau$) associated with this particular structures (see Tab. S1 given as ESI).\cite{cox2016quantum} Also, this time ZZ PRFs are redshifted with respect to AC ones. The plasmon modes associated to the main band are plotted in terms of the imaginary $\omega$FQ charges for both AC and ZZ configurations in Fig. \ref{fig:triangles}c. Also in this case plasmon modes have a dipolar character. 

Remarkably, all features of $\omega$FQ spectra are entirely confirmed by reference \emph{ab-initio} data \cite{cox2016quantum}(see Fig. \ref{fig:triangles}b, bottom). The agreement with our atomistic, yet classical, approach is impressive, and only a small blueshift of the $\omega$FQ PRFs with respect to their reference counterparts in noticed. However, such a discrepancy can be arbitrarily reduced by modifying $\omega$FQ parameters, which have been set to give the best agreement for all studied geometries on average (see Fig. S1 in the ESI). Similarly to nanoribbons, the electromagnetic simulations cannot reproduce the differences between AC and ZZ configurations, which are instead well described by $\omega$FQ. Finally, it is also worth noticing that $\omega$FQ correctly reproduces the degeneracy between $x-$ and $y-$ polarizations (see Fig. S5 in the ESI), a feature which has recently been reported for similar geometries \cite{myroshnychenko2018unveiling}. 

\begin{figure}[htbp!]
	\centering
	\includegraphics[width=1\linewidth]{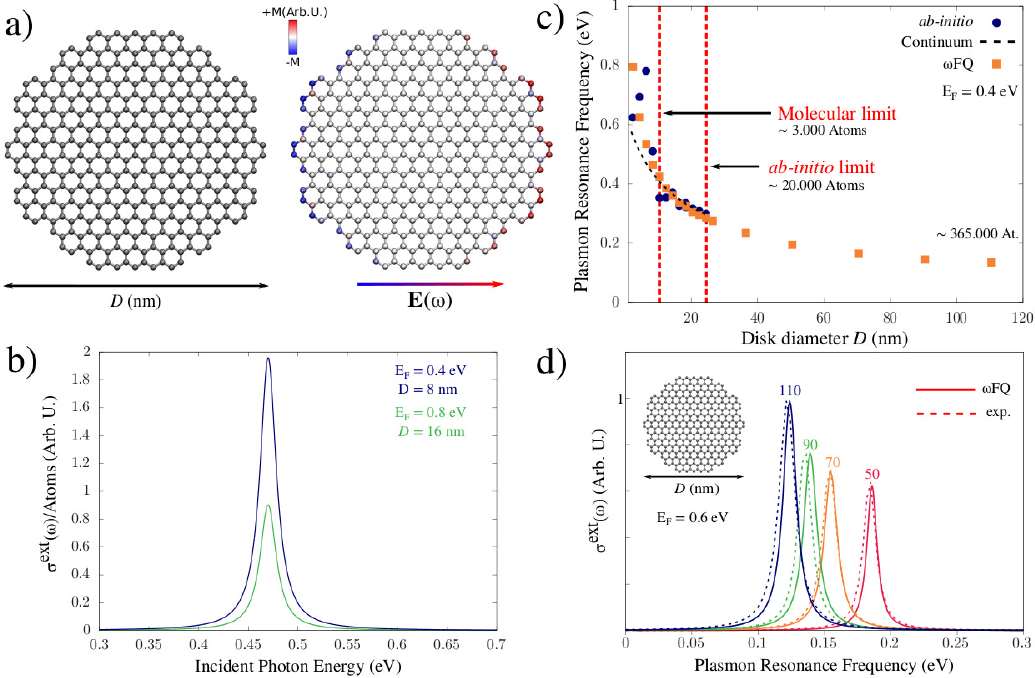}
	\caption{\textbf{(a)} Graphical depiction of the graphene disks studied in the present work. The main length scale $D$ (diameter length) is highlighted (left). Pictorial representation of $\omega$FQ imaginary charges representing the local plasmonic response of graphene disks (right). Colors are satured for $\pm$ 3.0$^{-1}$ a.u. The external electric field intensity is 10$^{-7}$ a.u. \textbf{(b)} $\omega$FQ $\sigma^{\text{ext}}$ of two graphene disks ($D$ = 8 nm and $D$ = 16 nm, respectively) calculated by imposing $\EF$ equal to 0.4 and 0.8 eV respectively. \textbf{(c)} $\omega$FQ (square orange points) and \emph{ab-initio} (Random Phase Approximation - RPA, circle points) Plasmon Resonance Frequency (eV) as a function of $D$ (from 2 to 110 nm, with a constant step of 2 nm). Classical continuum results are also depicted (dashed line). Both continuum and \emph{ab-initio} data are reproduced from Ref. \citenum{thongrattanasiri2012quantum}. E$_{\text{F}}$ is 0.4 eV in all calculations.  \textbf{(d)} $\omega$FQ $\sigma^{\text{ext}}$ of graphene disks with different diameter lengths (from 50 nm to 110 nm). The experimental data are reproduced from Ref. \citenum{fang2013gated}. E$_{\text{F}}$ is 0.6 eV. $\omega$FQ spectra are corrected by the factor $1/\sqrt{\epsilon}$, with $\epsilon = \dfrac{1+\varepsilon_{SiO_2}}{2}$, $\varepsilon_{SiO_2} = 2.30$.\cite{fang2013gated}}
	\label{fig:disks}
\end{figure}

As a third case study, $\omega$FQ is challenged against graphene-based disks of diameter $D$. Such nanodisks have been constructed by following the same procedure as described in Ref. \citenum{thongrattanasiri2012quantum} (see Fig. \ref{fig:disks}a, left). First, we selected two nanodisks with $D$ equal to 8 or 16 nm; their optical response was calculated by chosing $\EF =$ 0.4 and $\EF =$ 0.8 eV, respectively (see Fig. \ref{fig:disks}b). Both calculated $\omega$FQ spectra are characterized by an intense peak at about 0.47 eV. The plasmonic character of the associated plasmon is depicted in Fig. \ref{fig:disks}a, right, showing again the typical dipolar plasmon. The capability of $\omega$FQ to yield identical PRFs for the two studied systems confirms its robustness and reliability. In fact, the peculiar property of graphene-based materials to yield plasmon degeneracy as a result of a modulation by same numerical factor (in this case 2) of both the intrinsic dimensions of the considered substrate and the Fermi energy, is correctly reproduced \cite{yu2017analytical}. Also, the computed decreasing of the normalized extinction cross section $\sigma^{\text{ext}}$ (given by the sum of $\sigma^{\text{abs}}$ and the scattering cross section) is in perfect agreement with previous \emph{ab-initio} studies on similar graphene-based structures. \cite{yu2017analytical,thongrattanasiri2012quantum}

The dependence of the calculated $\omega$FQ PRF on the disk diameter is reported in Fig. \ref{fig:disks}c (orange squares) together with \emph{ab-initio} (blue circles) and continuum BEM (dashed line) data reproduced from Ref. \citenum{thongrattanasiri2012quantum}. \emph{Ab-initio} and BEM results are limited to disk diameters from 2 nm (112 atoms) to 24 nm (17272 atoms).\cite{thongrattanasiri2012quantum} The Fermi energy is fixed to 0.4 eV. For $D$ $>$ 14 nm, the $\omega$FQ PRF perfectly matches both \emph{ab-initio} and BEM data. In particular, $\omega$FQ and BEM values are almost identical for $D$ $>$ 8 nm, whereas the matching with \emph{ab-initio} values occurs for $D$ $>$ 14 nm. We notice that the small redshifts/blueshifts reported for some structures with $D$ $>$ 14 nm at the \emph{ab-initio} level with respect to BEM, are correctly reproduced by $\omega$FQ, thus confirming once again its capability to take into account the effects arising from small structural differences. When the diameter is lower than 8 nm, both \emph{ab-initio} and $\omega$FQ deviate from the BEM continuum curve; in particular a large blueshift arises, which is almost 0.2 eV for $D$ = 2 nm at $\omega$FQ level. However, for such small structures, the reference \emph{ab-initio} results do not present a clear trend as a function of the disk diameter, thus probably showing that the molecular limit is reached.\cite{thongrattanasiri2012quantum} As a consequence, $\omega$FQ cannot exactly reproduce the \emph{ab-initio} trend because the Drude model may fail in the limit of molecular excitations which are ruled by quantum mechanics.\cite{thongrattanasiri2012quantum,manjavacas2013plasmons} In Fig. \ref{fig:disks}c, $\omega$FQ PRF are also plotted for disks with $D$ $>$ 24 nm, which is the largest structure affordable by \emph{ab-initio} methods \cite{thongrattanasiri2012quantum}. Clearly, $\omega$FQ allows the calculation of structures that are more than 20 times greater than those affordable by state-of-the-art approaches.\cite{thongrattanasiri2012quantum,cox2016quantum} The plasmonic modes occurring in this range of dimensions are obviously well-described by classical approaches (such as BEM and $\omega$FQ). However, the atomistic nature of $\omega$FQ permits the investigation of complex geometrical arrangements, which can be hardly faced with purely continuum approaches.

To further demonstrate $\omega$FQ reliability and potentialities, in Fig. \ref{fig:disks}d we compare our results with experimental $\sigma^{\text{ext}}$ data measured for graphene-based disks with 50 nm $ < D <$ 110 nm.\cite{fang2013gated} Notice that such experimental measurements were conducted on graphene disks patterned on an ITO-coated silica substrate and covered with ion gel.\cite{fang2013gated} Notice that in $\omega$FQ, we have that $\sigma^{\text{abs}} \propto \dfrac{1}{\sqrt{\epsilon}}$, where $\epsilon$ is the relative permittivity constant of the surrounding environment.\cite{grigorenko2012graphene} Therefore, in order to match experimental conditions,\cite{fang2013gated} $\epsilon$ has been replaced by $\dfrac{1+\varepsilon_{SiO_2}}{2}$, where $\varepsilon_{SiO_2}$ is the permittivity constant of $SiO_2$ (i.e. 2.3).\cite{fang2013gated} By looking at Fig. \ref{fig:disks}d, the agreement between computed and experimental data is impressive and both peak relative positions and relative intensities perfectly match experimental values (with an error of about 0.01 eV). 

\begin{figure}[htbp!]
	\centering
	\includegraphics[width=1\linewidth]{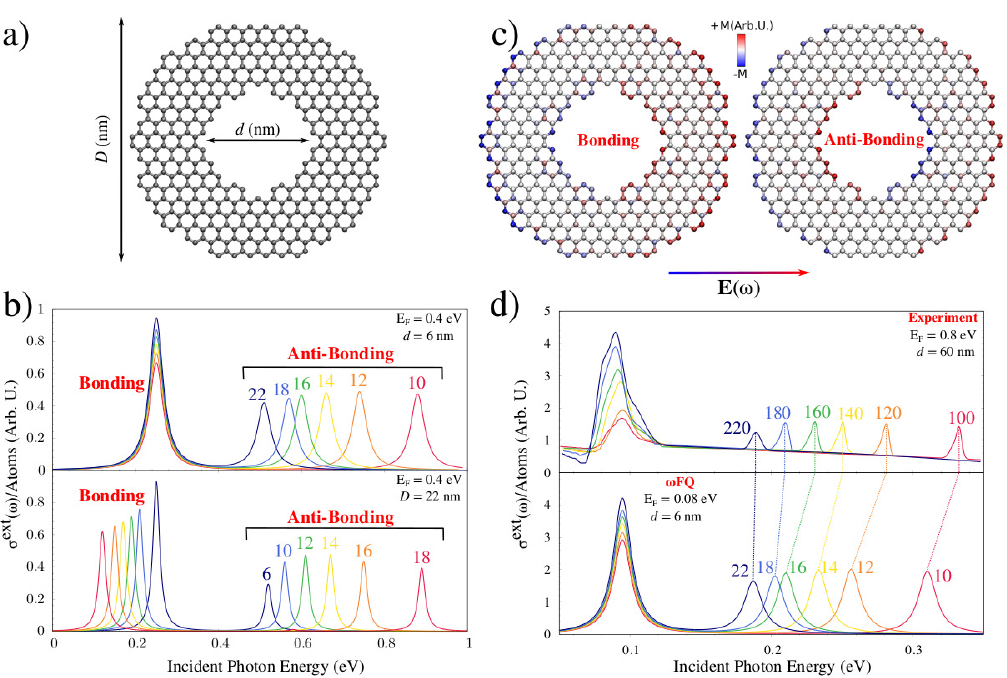}
	\caption{\textbf{(a)} Graphical depiction of graphene rings studied in the present work. The two relevant length scales $D$ (external diameter length) and $d$ (internal diameter length) are highlighted. \textbf{(b)} $\omega$FQ $\sigma^{\text{ext}}$ of graphene rings with fixed $d$ = 6 nm (top) and $D$ = 22 nm (bottom) as a function of $D$ (top) and $d$ (bottom). The length of the varied diameter is reported above each peak (in nm). Bonding and anti-Bonding plasmon modes are highlighted. The Fermi energy is 0.4 eV. \textbf{(c)} Pictorial representation of $\omega$FQ imaginary charges representing bonding (left) and anti-bonding (right) local plasmonic response for graphene rings. Colors are satured for $\pm$ 3.0$^{-1}$ a.u. The external electric field intensity is 10$^{-7}$ a.u. \textbf{(d)} Experimental\cite{fang2013gated} (top) and computed $\omega$FQ (bottom) $\sigma^{\text{ext}}$ of graphene rings with fixed $d$ = 60 nm (top) and $d$ = 6 nm (bottom) as a function of $D$ (from 220 to 100 nm -- top, from 22 to 10 nm -- bottom). The length of the varied diameter $D$ is reported above each peak (in nm). The experimental E$_{\text{F}}$ in the experiment is 0.8 eV (top), whereas $\omega$FQ E$_{\text{F}}$ is 0.08 eV (bottom). $\omega$FQ spectra are corrected by the factor $1/\sqrt{\epsilon}$, with $\epsilon = \dfrac{1+\varepsilon_{SiO_2}}{2}$, $\varepsilon_{SiO_2} = 2.30$.\cite{fang2013gated}}
	\label{fig:rings}
\end{figure}

As a last example, we applied $\omega$FQ to the calculation of the optical response properties of graphene-based rings, which are obtained by cutting an inner disk of diameter $d$ from a bigger disk of diameter $D$ (see Fig. \ref{fig:rings}a). Such a system was chosen to show that PRF can be tuned by modifying the internal diameter (d) and keeping fixed the external one ($D$) and viceversa. In fig. \ref{fig:disks}b, $\sigma^{\text{ext}}$ for a ring with $d$ equal to 6 nm (top) and with $D$ equal to 22 nm (bottom) was studied as a function of $D$ (top) and $d$ (bottom), respectively. In the former case, the spectrum is dominated by an intense peak at about 0.24 eV, whose PRF remains constant by changing $D$, and by a second band at higher energy which redshifts as $D$ increases (see Fig. \ref{fig:rings}b, top). In the second case (i.e. $D$ = 22 nm as a function of $d$, see Fig. \ref{fig:rings}b, bottom), the two bands are still present, but they show opposite trends as the dimension of the structure increases, i.e. the first blueshifts (PRF $<$ 0.3 eV) whereas the second reshifts (PRF $>$ 0.45 eV). In both cases, we studied the plasmonic nature of the two plasmon modes associated to the two bands by resorting to the so-called hybridization model,\cite{wang2006symmetry,wang2007plasmonic} which have been amply exploited to theoretically explain the plasmon excitations arising in structures presenting cavities \cite{prodan2003hybridization,bardhan2009nanosphere,park2009nature,radloff2004plasmonic,
wang2007plasmonic,wang2006symmetry,fang2013gated}. In particular, by plotting the imaginary $\omega$FQ charges calculated at the two PRFs we see that the typical bonding and anti-bonding modes, which are theoretically predicted by hybridization models, are perfectly described by $\omega$FQ. In Fig. \ref{fig:rings}b, the two bands are labeled and assigned to the two plasmonic modes. 
 
$\omega$FQ results are finally compared to the experimentally measured dependence of $\sigma^{\text{ext}}$ on $D$ for selected graphene-based rings with $d$ equal to 60 nm. $D$ varies from 100 nm to 220 nm and the experimental Fermi energy is set to 0.8 eV. Similar to the previous case (graphene disks), in the experimental measurements graphene rings were patterned on ITO-coated silica substrate and covered with ion gel.\cite{fang2013gated} Therefore, the same approach sketched above to correct $\sigma^{\text{abs}}$ has been exploited. To demonstrate the reliability of $\omega$FQ, we exploited the degeneracy property exposed above for nanodisks(see Fig. \ref{fig:disks}b). Therefore, we have multiplied by the same numerical factor both the intrinsic dimensions of the nanostructure ($d,D$) and the Fermi energy. In this particular case, such a degeneracy is obtained by dividing both the aforementioned quantities by 10, so that the studied rings are exactly the same as those discussed in Fig. \ref{fig:rings}b, but the Fermi energy this time is set to 0.08 eV. The agreement between the experimental\cite{fang2013gated} and computed $\omega$FQ extinction cross sections is particularly impressive, considering that all the most relevant experimental quantities (PRF of bonding and anti-bonding modes) are almost perfectly reproduced by $\omega$FQ (see Fig. \ref{fig:rings}d). Notice however that some discrepancies, in particular in case of the anti-bonding modes (above 0.2 eV), are present. These can be  due to  finite size effects of the computationally considered structures. As a last comment, we want to stress that $\omega$FQ can in principle afford experimental structures (which are constituted of $\sim 600.000$ atoms at most, for the largest structure). However, instead of showing the calculated data for the actual structures in this case we show that it is possible to hugely reduce the computational cost of the calculation, but keep the same accuracy with the actual structures, by taking advantage of the capability of $\omega$FQ to correctly model the aforementioned graphene physical features.


To conclude, in this work we have presented a novel classical, fully atomistic approach, which we dub ``$\omega$FQ'', to calculate the plasmonic properties of graphene-based nano- and micro-structures. Its potential and performance have been tested by comparing results obtained by exploiting this approach, against \emph{ab-initio}, continuum and experimental results.
Several shapes and dimensions have been taken into consideration, showing that, pending a reliable parametrization of the classical frequency-dependent force field, an almost perfect agreement with either reference \emph{ab-initio} or experimental data is achieved. In particular, the limitations of the state-of-the-art approaches, i.e. purely classical continuum and QM-based models, are completely overcome, because $\omega$FQ is able to treat realistic systems ($\sim$ 370.000 atoms) by retaining the atomistic picture of the studied structures and by showing at the same time a perfect agreement with experimental data. Simultaneously, for smaller graphene sheets, $\omega$FQ accuracy is perfectly in line with the best reference methods which have been presented in the literature.
In addition, the development of $\omega$FQ paves the way for an accurate description of the physico-chemical properties of molecules adsorbed on graphene-based substrates, thus allowing to deeply understand the nature of phenomena that have not been clearly explained, as for instance GERS. \cite{ling2009can,ling2010first,ling2013graphene,ling2012charge} Such an extension will require the coupling of $\omega$FQ with a Quantum Mechanical (QM) description of the adsorbed molecule, in a QM/Molecular Mechanics (QM/MM) fashion.\cite{senn2009qm,morton2011theoretical,payton2013hybrid,giovannini2019fqfmu,
rinkevicius2014hybrid,cappelli2016integrated,giovannini2019fqfmuder2} Such a development will be the topic of future publications.

\section*{Methods}

The $\omega$FQ approach has been implemented in a stand alone Fortran 95 package. Eq. \ref{eq:wFQ} is solved for a set of frequencies given as input. In particular, all computed spectra reported in the manuscript were obtained by explicitly solving linear response equations for steps of 0.01 eV. The final quantity that is obtained from solving Eq. \ref{eq:wFQ} are complex $\omega$FQ charges $q$, which are then used to define the complex polarizability $\overline{\alpha}$ as:

\begin{equation}
\label{eq:polar}
\overline{\alpha}(\omega)_{kl} = \parz{\overline{\mu}_k(\omega)}{E_l(\omega)} = \sum_i q_i(\omega) \cdot \frac{k_i}{E_l(\omega)}
\end{equation}
where $\omega$ is the frequency of the incident field, which has an intensity $E(\omega)$. $\bm{\mu}$ is the complex dipole moment, $i$ runs over graphene atoms, $k$ represents $x,y,z$ positions of the $i$-th atom, and $l$ runs over $x$,$y$,$z$ directions.  

Finally, the absorption ($\sigma^{\text{abs}}$), the scattering ($\sigma^{\text{sca}}$) and the extinction ($\sigma^{\text{ext}}$) cross sections can be calculated:

\begin{align}
\label{eq:css_sec}
\sigma^{\text{abs}} & = \frac{4\pi}{3c} \omega \ \text{tr} \bigl(\alpha^* \bigr) \nonumber \\
\sigma^{\text{sca}} & = \frac{8\pi}{3c^4} \omega^4 \ \left[\text{tr} \bigl(\alpha \bigr)^2 + \text{tr} \bigl(\alpha^* \bigr)^2 \right] \nonumber \\
\sigma^{\text{ext}} & = \sigma^{\text{abs}} + \sigma^{\text{sca}}
\end{align}

where $\alpha$ and $\alpha^*$ are the real and the imaginary part of the complex polarizability $\overline{\alpha}$, respectively.

For all the studied graphene nanostructures, the parameters exploited in Eq. \ref{eq:wFQ} were extracted from physical quantities recovered from the literature or numerically tested on selected systems (see ESI for more details). The parameters finally exploited are the following (see ESI for more details): $\tau = 4.15\cdot 10^{-13}$ s,\cite{thongrattanasiri2012quantum} $\alpha=0.0031$, $A_{ij} = 0.4879\cdot 10^{-20}$ m$^2$, $l^0_{ij} = 1.42\cdot 10^{-10}$ m,\cite{neto2009electronic} $d = 12.00$,\cite{giovannini2019classical} $s = 1.10$.\cite{giovannini2019classical}

\section*{Electronic Supplementary Information}

Detailed derivation of the $\omega$FQ model for 2D substrates. Model parametrization. Structural details of the studied systems. Validation of $\omega$FQ model.



\section*{Conflicts of Interest}

There are no conflicts of interest to declare.

\section*{Acknowledgments}

This work has received funding from the European Research Council (ERC) under the European Union’s Horizon 2020 research and innovation programme (grant agreement No. 818064). TG acknowledges funding from the Research Council of Norway through its grant TheoLight (grant no. 275506). MP was supported by the European Union's Horizon 2020 research and innovation programme under grant agreement No.~785219 - GrapheneCore2. 



\newpage

\newpage

\newpage

\newpage

\cleardoublepage

\newpage

\small
\bibliographystyle{rsc}
\bibliography{biblio}

\end{document}